\documentclass[amssymb,prb,preprint,aps,showpacs]{revtex4}
\usepackage{epsfig}
\usepackage{mathrsfs}
\begin{document}

\title{\textbf{Regularization of fluctuations near the sonic
horizon due to the quantum potential and its influence on the
Hawking radiation}}
\author{V. Fleurov$^{1,3}$, R. Schilling$^2$}
\affiliation{
$^{1}$  Raymond and Beverly Sackler Faculty of Exact Sciences,\\
School of Physics and Astronomy, Tel-Aviv University, Tel-Aviv
69978, Israel
\\
$^2$ Johannes Gutenberg University, Mainz, Germany
\\
$^{3}$ Max-Planck-Institut f\"ur Physik Komplexer Systeme,
N\"othnitzer Strasse 38, 01187 Dresden, Germany}

\begin{abstract}
We consider dynamics of fluctuations in transonically accelerating
Bose-Einstein condensates and/or luminous fluids (coherent light
propagating in a Kerr nonlinear medium) using the hydrodynamic
approach. It is known that neglecting the quantum potential (QP)
leads to a singular behavior of quantum and classical fluctuations
in the vicinity of the Mach (sonic) horizon, which in turn gives
rise to the Hawking radiation. The neglect of QP is well founded at
not too small distances $|x| \gg l_h$ from the horizon, where $l_h$
is the healing length. Taking the QP into account we show that a
second characteristic length $l_r > l_h$ exists, such that the
linear fluctuation modes become regularized for $|x| \ll l_r$. At
$|x| \gg l_r$ the modes keep their singular behavior, which however
is influenced by the QP. As a result we find a deviation of the high
frequency tail of the spectrum of Hawking radiation from Planck's
black body radiation distribution, which can be described by an
effective Hawking temperature decreasing with increasing frequency.
Similar results hold for the wave propagation in Kerr nonlinear
media where the lengths $l_h$ and $l_r$ exist due to the
nonlinearity.
\end{abstract}

 \maketitle

\section{Introduction}

Hawking radiation is one of the most impressive phenomena at the
intersection of general relativity and quantum field theory. The
account of quantum nature of the physical vacuum in curved space led
to the prediction that a black hole - an object defined classically
as an object that even light cannot escape - in fact emits
radiation,\cite{H75,H76} which is characterized by the Planck's
distribution of the black body at a certain temperature. The
derivation involves singular behavior for the eigenmodes of the
proper field equation in the vicinity of the horizon, which can be
normalized only by using assumptions on Planck scale physics that
cannot be justified completely. For a general discussion the reader
may consult Ref. \onlinecite{BD84}.

Soon after that development it was suggested\cite{U81} to consider
analogous phenomena in condensed matter physics where the
"high-energy" (short-wavelength) physics is known. The suggestion
was based on the observation that the derivation of the Hawking
radiation uses only the linear wave equation in curved space-time,
and does not make the actual use of the Einstein equations for
gravity. The same conditions for wave propagation arise at the
consideration of sound propagation in the fluid when the background
flow is non-trivial.\cite{U81} In particular, the background may be
that of a stationary accelerating transonic flow where a surface
appears, on which the fluid velocity equals that of sound, i.e where
the Mach number is one, $M = 1$. This surface has many features
similar to that of the black hole horizon and is often called sonic
or Mach horizon.

Currently theoretical studies of such artificial black holes became
a very active field involving a variety of physical
systems.\cite{JV98,R00,G05,BLV03,CFRBF08,RPC09,NBRB09,FFBF10}
Experimentally a white-hole horizon was observed in optical
fibers,\cite{PKRHK08} where the probe light was back-reflected from
a moving soliton, and a black-hole horizon was observed in a
Bose-Einstein condensate (BEC) system.\cite{LIBGRZS10} Continuing
the study of optical fibers a radiation has been quite recently
observed,\cite{BCCGORRSF10} which is a very promising contender to
being analogous to Hawking radiation in a table-top experiment. A
"horizon physics" is currently intensively studied also in the
surface water waves. \cite{RMMPL08,RMMPL10,WTPUL11}

\begin{figure}[tbp!]
\begin{center}
\includegraphics[width=0.95\columnwidth]{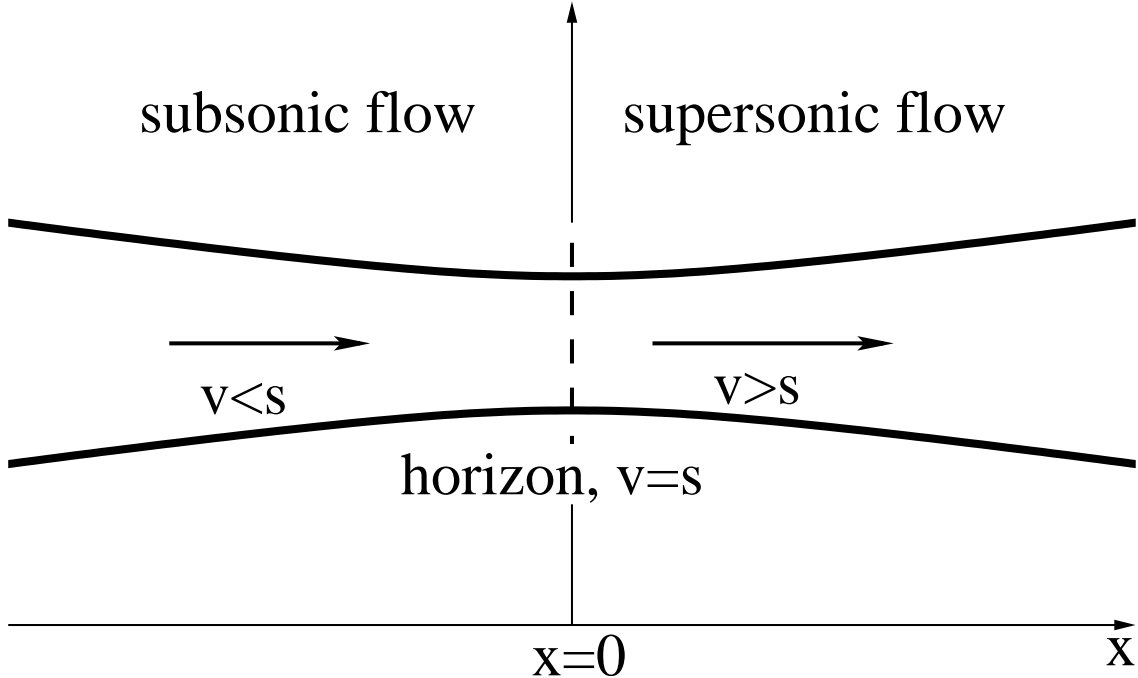}
\end{center}
\caption{A scheme of Laval nozzle as a vessel with non-monotonously
varying cross-section. It is capable of accelerating a flow from
subsonic to supersonic velocity. The sound velocity is reached in
the throat (the narrowest cross-section) as shown by the dashed line
where a Mach (sonic) horizon is formed.} \label{Laval.fig}
\end{figure}

Creating a stationary laminar transonic flow of an ordinary fluid,
as required by the original proposal by Unruh,\cite{U81} may be a
formidable task that is why many authors turn their attention to the
BEC (see, e.g. Refs. \onlinecite{BLV03,CFRBF08,RPC09}), or as
recently proposed, to a "luminous fluid"\cite{FFBF10} in an optical
analog of the Laval nozzle. Generally the Laval nozzle (see, e.g.,
Refs. \onlinecite{S65,LL87}) schematically shown in Fig.
\ref{Laval.fig} accelerates a flow from sub- to supersonic
velocities. It uses the fact that the flow accelerates if the
cross-section of the vessel decreases in the subsonic region,
whereas in the supersonic region the cross-section must increase in
order to accelerate the flow. An optical analog of the Laval
nozzle\cite{FFBF10} is a properly shaped wave guide filled with a
Kerr nonlinear medium, in which a coherent laser light propagates.

BEC is usually described by the Gross-Pitaevskii equation, which can
be converted into two equivalent hydrodynamic equations by means of
the Madelung transformation.\cite{M27} A similar procedure exists
also for the coherent light.\cite{M75} Analysis of the behavior of
fluctuations near the Mach horizon based on these equations is
usually carried out neglecting the role of the quantum potential
(QP).\cite{BLV03,FFBF10} This approximation can be well justified
nearly everywhere except for the narrow vicinity of the sonic
horizon due to the singular behavior of the fluctuations in this
region. Neglecting QP the fluctuations are described by an
equivalent Klein-Gordon equation for the phase fluctuations in a
curved space with Schwarzschild metric, as first proposed in Ref.
\onlinecite{U81}. The reader should also note that the fluctuation
modes are then dispersionless, i.e. their frequency depends linearly
on their wave-number, if QP is neglected.

At small distances from the horizon the contribution of the QP
(kinetic energy operator for atoms in BEC or dispersion in luminous
fluid) diverges and cannot be anymore neglected. This reminds the
behavior of vacuum fluctuations near the black hole horizon on the
Planck scale. Since in our case the physics on small scales is known
the role of the QP in the near vicinity of the Mach horizon can and
should be addressed.

Accounting for the QP introduces dispersion effects at higher
frequencies corresponding to small length scales. The question how
far such effects influence Hawking radiation has been studied during
the last two decades. It seems that the first such study was done by
Jacobson \cite{X1,X2}. Whereas Ref. \onlinecite{X1} had stated that
a derivation of black hole radiation, taking the ultrashort distance
physics into account, was missing, such a derivation was presented
in Ref. \onlinecite{X2}. It has been concluded that boundary
conditions can be formulated, which lead to the Hawking effect. In
order to deduce quantitative results, e.g. for the flux and the
spectrum, the mode equations based on specific dispersion laws were
solved numerically \cite{X3,X4,X5a,X5b,RPC09} and analytically.
\cite{X6,X7,X8,X9,LKO03a,LKO03b,X12,X13,X15} In all of these papers
clear evidence has been given that Hawking radiation is basically a
low frequency phenomenon and is robust, although counterexamples
were presented\cite{X12} predicting strong deviation from the
Hawking's result. Due to dispersion an upper threshold
$\omega_{max}$ exists in the superluminal case such that the Hawking
radiation disappears for $\omega > \omega_{max}$.\cite{X5a,X5b} For
more details the reader is also referred to the review in Ref.
\onlinecite{X16}.

In the current paper we will focus on BEC or equivalently on the
light in a Kerr nonlinear medium. Assuming a stationary background
solution with inhomogeneous density, $\rho({\bf r})$, and velocity,
${\bf v}({\bf r})$, fields we will derive the linearized equation
for the fluctuations of both fields, $\delta\rho({\bf r})$ and
$\delta {\bf v}({\bf r})$, accounting for the QP. It will have a
form more complicated than the Klein-Gordon equation of Ref.
\onlinecite{U81}. Reducing the problem to a spatially one
dimensional one, two Hermitian fields $\chi(x,t)$ and $\xi(x,t)$ are
introduced, which obey two partial differential equations. We show
that assuming {\em linear} dependence of the background velocity and
density on the coordinate near the horizon $x = 0$, the equations
can be decoupled in the momentum space, and reduced to an equation
for $\chi_k$ representing the density fluctuations $\delta\rho({\bf
r})$. Our mathematical analysis is close to that of Ref.
\onlinecite{X8}, which makes use of the Laplace back transform.
However, there are several formal and physical points of difference
from Ref. \onlinecite{X8}, which will be discussed below. In
addition the resulting dispersion relation is dictated by the GP
(NLS) equation (\ref{GP}) (see next section) with the coordinate
dependent background flow  rather than chosen by the mathematical
requirement of nonlinearity \cite{X4,X5a,X5b,X7,X8,X9} or assuming a
rather general nonlinear dispersion.\cite{X3,X12,X13}

\section{Equations for BEC and luminous fluids}

We start by presenting a brief overview of the basic equations
controlling the dynamics of BEC or light in Kerr medium. This will
allow us to make necessary definitions and introduce basic notions
needed for the further analysis. Dynamics of Bose-Einstein
condensate is quite successfully described by the Gross - Pitaevskii
(GP) equation
\begin{widetext}
\begin{equation}\label{GP}
i\hbar \partial_t \Phi({\bf r}, t) = - \frac{\hbar^2}{2m} \nabla^2
\Phi({\bf r}, t) + U_{ext}({\bf r}) \Phi({\bf r}, t) + g |\Phi({\bf
r}, t)|^2 \Phi({\bf r}, t)
\end{equation}
\end{widetext}
where $\Phi({\bf r},t)$ is the condensate wave function of identical
Bose particles with the mass $m$, $U_{ext}({\bf r})$ is an external
potential, and $g$ is the parameter of the particle interaction. The
equation is usually derived (see, e.g. Ref. \onlinecite{FV71}) from
the equation of motion for the boson field operator
$\widehat{\Psi}({\bf r}, t)$, which is represented in the form
\begin{equation}\label{fluctuations-a}
\widehat{\Psi}({\bf r}, t) = \Phi({\bf r}, t) + \widehat{\psi}({\bf
r}, t)
\end{equation}
so that the field operator $\widehat{\psi}({\bf r}, t)$ describes
fluctuations near the condensate state $\Phi({\bf r})$. For small
fluctuations the corresponding linearized equation reads
\begin{widetext}
\begin{equation}\label{linear-a}
i\hbar \partial_t \widehat{\psi}({\bf r},t) = \left(-
\frac{\hbar^2}{2m} \nabla^2 + U_{ext}({\bf r}) - \mu\right)
\widehat{\psi}({\bf r},t) + 2g |\Phi({\bf r})|^2 \widehat{\psi}({\bf
r},t) + g \Phi^2({\bf r}) \widehat{\psi}^\dagger({\bf r},t)
\end{equation}
\end{widetext}
Usually solution $\Phi({\bf r})$ of the stationary GP equation is
used, which is obtained from Eq. (\ref{GP}) by replacing its left
hand side by $\mu \Phi({\bf r})$, where $\mu$ is the chemical
potential.

A similar mathematical description is applied to the coherent light
propagating in Kerr nonlinear medium (see, e.g., Ref.
\onlinecite{A95}). The corresponding equation
\begin{equation}\label{NLS}
i \partial_z A =
-\frac{1}{2 \beta_0}\widetilde{\nabla}^2A -
\frac{\omega_0^2}{2\beta_0c^2} \Delta n^2(x,y) A + \lambda |A|^2 A.
\end{equation}
is usually called Nonlinear Schr\"odinger (NLS) equation. The latter
is derived from the classical Maxwell equations for electromagnetic
waves in a Kerr nonlinear medium. The paraxial approximation is
assumed, according to which the electric component of the wave is
represented in the form
$$
E = \int \frac{d\omega}{2\pi}A(x,y;z;\omega) e^{-i\beta_0 z}
e^{i\omega t}
$$
where the amplitude $A$ weakly depends on the coordinate $z$ along
the light propagation axis. As for its frequency dependence it is
narrowly peaked around the principal frequency $\omega_0$. Here
$\beta_0 = k(\omega_0)$ where $k^2(\omega) = \omega^2\mu_0
\varepsilon(\omega)$. The magnetic susceptibility $\mu_0$ is assumed
to be a constant, whereas the frequency dependence of the dielectric
function $\varepsilon(\omega)$ plays an important role. $c$ is the
light velocity. Usually the light polarization does not change,
which allows one to treat this equation as a scalar one.

Although NLS Eq.(\ref{NLS}) is classical we can formally multiply it
by $\hbar$ and get the correspondence to the quantum GP equation
where the propagation distance $z$ plays the part of time, $\hbar
\beta_0$ is the mass of a fictitious particle, which is measured in
units of momentum. It is interesting to note that a quantity
corresponding to energy or chemical potential should be also
measured in units of momentum. It implies also that the "velocity"
is dimensionless. Spacial variation $U_{ext}(x,y) = - \hbar
\frac{\omega_0^2}{2\beta_0c^2} \Delta n^2(x,y)$ of the refraction
index $n^2 = \mu_0 \varepsilon(\omega) c^2$ plays now the part
analogous to that of the external potential in the GP equation and
the field dependence of the dielectric function results in the
nonlinear term in Eq.(\ref{NLS}).

The time coordinate appears now as one of the "space" coordinates in
the effective Laplacian
\begin{equation}\label{eff-Laplacian}
\widetilde{\nabla}^2 = \partial^2_x + \partial^2_y \pm
\partial^2_\tau
\end{equation}
where
$$
\tau = \sqrt{\frac{1}{2\pi |D| \beta_0}}\left(t - \frac{z}{v_g}
\right),
$$
$v_g^{-1} = \left.\frac{\partial k(\omega)}{\partial\omega}
\right|_{\omega=\omega_0}$ is the light group velocity in the medium
and $D = \frac{1}{2\pi} \left.\frac{\partial^2 k(\omega)}{\partial
\omega^2}\right|_{\omega = \omega_0}$ is the dispersion coefficient.
The minus sign in Eq. (\ref{eff-Laplacian}) corresponds to the
normal dispersion ($D > 0$) whereas the plus sign corresponds to the
anomalous dispersion ($D < 0$). The hyperbolic "Laplacian" for the
normal dispersion can often occur for the coherent light propagating
in a Kerr medium but does not appear in the GP equation for BEC
gases. We can always choose the frequency window within the
anomalous dispersion regime such that the Laplacian keeps its
standard shape for the coherent light as well. When a laser emits a
pulse, $z$ is actually the position of the pulse reached during the
time $z/v_g$. $\tau$ is used to describe the shape of the pulse (in
the $z$ direction) in the coordinate system moving together with it.

The amplitude $A$ can be always rescaled in such a way as to make
its dimension coincide with that of the wave function $\Phi$. It
does not influence the linear part of Eq.(\ref{NLS}), whereas the
factor appearing in the nonlinear term can be absorbed into the
parameter $\lambda$, which together with $\hbar$ then maps onto $g$
of Eq.(\ref{GP}), making the mapping between the two equations
nearly complete.

Now we can also consider stationary solution of the NLS equation and
small fluctuations around it,
\begin{equation}\label{fluctuations-b}
A(\widetilde{{\bf r}}, z) = \Phi(\widetilde{{\bf r}}) +
\psi(\widetilde{{\bf r}}, z).
\end{equation}
where $\widetilde{{\bf r}} = (x,y,\tau)$. $\psi(\widetilde{{\bf r}},
z)$ satisfies an equation, which maps onto Eq. (\ref{linear-a})
according to the above prescriptions with the only difference that
$\psi$ in Eq.(\ref{fluctuations-b}) is a scalar function rather than
a field operator of Eq.(\ref{fluctuations-a}).

\section{Equation of motion for fluctuations}

The hydrodynamic description of the Heisenberg equation of motion
for the field operators $\widehat{\Psi}({\bf r},t)$ and
$\widehat{\Psi}^\dagger({\bf r},t)$ follows by introducing the
density operator
$$
\widehat{\rho}({\bf r},t) = m \widehat{\Psi}^\dagger({\bf
r},t)\widehat{\Psi}({\bf r},t)
$$
and current density operator
$$
\widehat{{\bf j}}({\bf r},t) =
\frac{1}{2i}[\widehat{\Psi}^\dagger({\bf r},t) \nabla
\widehat{\Psi}({\bf r},t) - (\nabla\widehat{\Psi}^\dagger({\bf
r},t))\widehat{\Psi}({\bf r},t)].
$$
Both are connected by the continuity equation
$$
\partial_t \widehat{\rho}({\bf r},t) + \nabla \widehat{{\bf j}}({\bf
r},t) = 0.
$$

Substituting Eq.(\ref{fluctuations-a}) into Heisenberg equations of
motion for $\widehat{\Psi}({\bf r},t)$ and
$\widehat{\Psi}^\dagger({\bf r},t)$, using $\Phi({\bf r},t) = f({\bf
r},t) \exp[-i\varphi({\bf r},t)]$ and neglecting fluctuations one
obtains hydrodynamic equations
\begin{equation}\label{continuity}
\frac{\partial}{\partial t} \rho + \nabla\cdot [\rho{\bf v}] = 0,
\end{equation}
\begin{equation}\label{Euler}
\frac{\partial}{\partial t} {\bf v} + \frac{1}{2}\nabla {\bf v}^2 =
- \frac{1}{m} \nabla \left[U_{qu} + U_{ext} + g f^2 \right]
\end{equation}
for the density $\rho({\bf r},t) = m f^2({\bf r},t)$ and velocity
${\bf v}({\bf r},t) = - \frac{\hbar}{m}\nabla \varphi({\bf r},t) $;
both $f$ and $\varphi$ are real functions. These two equations are
equivalent to the GP equation (\ref{GP}). Here
\begin{equation}\label{QP}
U_{qu} = - \frac{\hbar^2}{2m} \frac{\nabla^2 f}{f}
\end{equation}
is QP. This hydrodynamic representation of the basic equations
allows us to think about the BEC fluid or luminous fluid,
respectively.

We now consider small fluctuations of these two fields with respect
to a stationary solution $\Phi_0({\bf r}) = f_0({\bf r})
\exp[-i\varphi_0({\bf r})]$. It follows from Eq.(\ref{continuity})
that
\begin{equation}\label{continuity-st}
\nabla\cdot [\rho_0({\bf r}){\bf v}_0({\bf r})] = 0,
\end{equation}
where $\rho_0({\bf r}) = m f_0^2({\bf r})$ and ${\bf v}_0({\bf r}) =
- \frac{\hbar}{m}\nabla \varphi_0({\bf r}) $. To describe the
fluctuations we introduce the dimensionless quantities
\begin{equation}\label{operators}
\begin{array}{c}
\chi = \frac{1}{f_0} \left[ e^{-i\varphi_0} \psi^\dagger +
e^{i\varphi_0} \psi \right],
\\
\xi = \frac{1}{2if_0}[e^{-i\varphi_0} \psi^\dagger - e^{i\varphi_0}
\psi],
\end{array}
\end{equation}
which are field operators in the case of BEC or classical scalar
fields in the case of coherent light. Therefore we skipped the hats.
These quantities are connected with the density and velocity
fluctuations by the relations
\begin{equation}\label{fluctuations-c}
\begin{array}{c}
\delta {\bf v}({\bf r},t) = - \displaystyle \frac{\hbar}{m}
\nabla\xi({\bf r},t),
\\
\delta\rho({\bf r},t) = \rho_0({\bf r}) \chi({\bf r},t).
\end{array}
\end{equation}
Then it is straightforward to show that Eq.(\ref{fluctuations-a})
and its Hermitian conjugate lead to
\begin{equation}\label{linear-b}
\begin{array}{c}
\widehat{D} \chi - \displaystyle \frac{\hbar}{m}\frac{1}{f_0^2}
\nabla(f_0^2 \nabla \xi) = 0,
\\ \\
\displaystyle \widehat{D} \xi + \frac{\hbar}{4m}
\frac{1}{f_0^2}\nabla( f_0^2 \nabla \chi) - \frac{1}{\hbar}gf_0^2
\chi =0.
\end{array}
\end{equation}
where $\widehat{D} = \partial_t + {\bf v}_0 \cdot \nabla $. These
equations also follow from Eqs.(\ref{continuity}) and (\ref{Euler})
in the linear order in $\delta\rho$ and $\delta{\bf v}$. While
deriving equations (\ref{linear-b}), we used
Eq.(\ref{continuity-st}).

It is instructive to indicate that Eqs. (\ref{linear-b}) are Euler
-- Lagrange equations (see also discussion in Appendix of Ref.
\onlinecite{X5b}) corresponding to the Lagrangian density
\begin{widetext}
\begin{equation}\label{Lagrangian-a}
{\cal L} = \frac{1}{2} f_0^2(\chi\dot\xi - \xi\dot\chi) +
\frac{1}{2} f_0^2 {\bf v}_0\cdot (\chi \nabla \xi - \xi \nabla \chi)
- \frac{1}{2\hbar}g f_0^4 \chi^2 - \frac{\hbar}{2m} f_0^2
(\nabla\xi)^2 - \frac{\hbar}{8m} f_0^2 (\nabla\chi)^2
\end{equation}
\end{widetext}
This Lagrangian density allows us also to define canonical momenta
\begin{equation}\label{canonicalmomentum}
p_\chi = \frac{\delta}{\delta \dot\chi} \left[ {\cal L} -
\frac{1}{2}\partial_t(f_0 \chi\xi) \right] = - f_0^2 \xi,\ \ \ p_\xi
= \frac{\delta}{\delta \dot\xi} \left[ {\cal L} +
\frac{1}{2}\partial_t(f_0 \chi\xi) \right] =  f_0^2 \chi
\end{equation}
for each of the fields $\chi$ and $\xi$. Together they form two
canonical pairs, each of which can be used to derive Hamiltonian.
The Hamiltonian density for these two cases can be derived as
\begin{equation}\label{hamiltonian}
{\cal H} = -f_0 \dot\chi\xi - \left[ {\cal L} -
\frac{1}{2}\partial_t(f_0 \chi\xi) \right] = f_0 \chi \dot\xi -
\left[ {\cal L} + \frac{1}{2}\partial_t(f_0 \chi\xi) \right] =
$$$$
- \frac{1}{2} f_0^2 {\bf v}_0\cdot (\chi \nabla \xi - \xi \nabla
\chi) + \frac{1}{2\hbar}g f_0^4 \chi^2 + \frac{\hbar}{2m} f_0^2
(\nabla\xi)^2 + \frac{\hbar}{8m} f_0^2 (\nabla\chi)^2.
\end{equation}
Although the derivation (\ref{hamiltonian}) depends on the choice of
the canonical pair, the final Hamiltonian does not depend on it.
Eqs. (\ref{linear-b}) are now just the pair of Hamilton equations of
motion for the Hamiltonian density (\ref{hamiltonian}). The above
manipulations with adding or subtracting a full derivative indicate
that the variations of the Lagrangian with respect to $\dot\chi$ and
$\dot\xi$ are not uniquely defined. This feature of the Lagrangian
(\ref{Lagrangian-a}) does not however effect the Euler- Lagrange
equations (\ref{linear-b}) but is important when defining the
canonical momentum and Hamiltonian. It means that we have to choose
which of the functions $\chi$ or $\xi$ plays the part of the
canonical coordinate, then the other one will enter the definition
of the conjugate canonical momentum. It is emphasized that we can
use only one of these pairs at a time, but not both of them
simultaneously. The physical result is independent of the choice.

If $\xi$ is chosen as the canonical coordinate then the quantization
condition reads
\begin{equation}\label{quantization}
[p_\xi({\bf r},t), \xi({\bf r}',t)] = [f_0^2 \chi({\bf r},t),
\xi({\bf r}',t)] = \hbar \delta({\bf r} -{\bf r}')
\end{equation}
where $[\cdots,\cdots]$ denotes the commutator of two operators. The
other choice of the canonical pair would obviously yield the same
quantization condition. One can readily see that the quantization
condition (\ref{quantization}) in the BEC case is consistent with
the definition (\ref{operators}), where $\psi^\dagger$ and $\psi$
are Bose field operators. In case of the coherent light when these
are scalar functions, Eq. (\ref{quantization}) allows us to quantize
them thus removing the last remaining difference between BEC and
luminous liquid.

We may extend the definition of the fields $\chi$ and $\xi$ and
consider them complex. Then the Lagrangian density
(\ref{Lagrangian-a}) can be represented in the form
\begin{widetext}
\begin{equation}\label{Lagrangian-b}
{\cal L} = \frac{i}{4} f_0^2 [\vartheta^\dagger \sigma_y
(\widehat{D}{\vartheta}) - (\widehat{D} \vartheta^\dagger) \sigma_y
{\vartheta}] - \frac{1}{2\hbar}g f_0^4 \vartheta^\dagger (1 +
\sigma_z) \vartheta - \frac{\hbar}{4m} f_0^2
(\nabla\vartheta^\dagger) \nabla\vartheta
\end{equation}
\end{widetext}
where the two component field
\begin{equation}\label{two-component}
\vartheta = \left(\begin{array}{c}
\frac{1}{\sqrt{2}}\chi \\
\sqrt{2} \xi
\end{array}
\right)
\end{equation}
has been introduced.

The corresponding equation of motion reads
\begin{widetext}
\begin{equation}\label{eq-motion-a}
\frac{i}{2} f_0^2 \sigma_y\dot{\vartheta} + \frac{i}{2} f_0^2 {\bf
v}_0 \cdot \sigma_y\nabla \vartheta - \frac{1}{2\hbar}g f_0^4 (1 +
\sigma_z) \vartheta + \frac{\hbar}{4m} \nabla (f_0^2
\nabla\vartheta) = 0.
\end{equation}
\end{widetext}
It is obviously equivalent to two Eqs.(\ref{linear-b}). Now we
multiply Eq.(\ref{eq-motion-a}) by $2 \vartheta^\dagger$ from the
left, write also a Hermitian conjugate equation and take their
difference. The result is
\begin{widetext}
$$
\partial_t(f_0^2 \vartheta^\dagger\sigma_y\vartheta) + \nabla
(f_0^2{\bf v}_0 \vartheta^\dagger \sigma_y \vartheta ) -
\frac{i\hbar}{2m} \nabla[f_0^2 [ \vartheta^\dagger \nabla\vartheta -
(\nabla\vartheta^\dagger) \vartheta ]] = 0.
$$
\end{widetext}
We may conclude that the quantity $\varrho = f_0^2
\vartheta^\dagger\sigma_y\vartheta$ plays the role of the density of
this two component field, whereas
\begin{equation}\label{current}
{\bf j} = f_0^2{\bf v}_0 \vartheta^\dagger \sigma_y \vartheta  - i
\frac{\hbar}{2m} f_0^2 [ \vartheta^\dagger \nabla\vartheta -
(\nabla\vartheta^\dagger) \vartheta ]
\end{equation}
is the corresponding current. The integral
\begin{equation}\label{KGscalar}
\int d^3r f_0^2 \vartheta^\dagger\sigma_y\vartheta = -i \int d^3r
f_0^2 (\chi^*\xi - \xi^* \chi)
\end{equation}
is a conserved quantity, which can be used in order to normalize
solutions (eigenfunctions) of Eq.(\ref{linear-b}) or to define a
scalar product of two such solutions. Similarly to the well-known
Klein-Gordon norm, Eq.(\ref{KGscalar}) is not positively defined.
Positive and negative frequencies are considered separately in order
to resolve this problem in the case of the Klein - Gordon equation
(see, e.g. Ref. \onlinecite{BD84}). A similar procedure will be
applied in this case as well. This property will be important for
the further discussion.

\section{Eigenfunctions}

\subsection{Bogoliubov spectrum}\label{bogoliubov.sec}

As an example we first may consider a homogeneous background
solution $\Phi({\bf r})$, in which only the phase may linearly
depend on the coordinate, i.e. both the density $\rho_0 = m f_0^2$
and the velocity $m {\bf v}_0 = - \hbar \nabla\varphi_0$ are
constants. Then by solving the second equation in
Eq.(\ref{linear-b}) for $\chi$ and substituting the result in the
first one this system reduces to a linear partial differential
equation with constant coefficients:
\begin{equation}\label{linear-e}
\displaystyle \widehat{D}^2 \xi - \frac{g}{m^2} \rho_0 \nabla^2 \xi
+ \frac{1}{\hbar^2} \left(\frac{\hbar^2}{2m }\nabla^2 \right)^2 \xi
= 0.
\end{equation}
It straightforwardly leads to the celebrated Bogoliubov spectrum
\begin{equation}\label{bogoliubov}
[\omega - {\bf v}_0\cdot{\bf k}]^2 = \overline{s}^2 k^2 \left[1 +
\frac{k^2 l_h^2}{2}\right] \equiv \Omega^2(k)
\end{equation}
in the coordinate system moving with the constant velocity ${\bf
v}_0$. The quantity $\overline{s}\ ^2 = \frac{g\rho_0}{m^2}$ is
obviously the sound velocity in the long wave limit and $ l_h^2 =
\hbar^2/(2mgf_0^2)= \hbar^2/(2m^2\overline{s}\ ^2)$ is the healing
length in BEC and the nonlinearity length $l^2_h =
1/(2\beta_0\lambda|A|^2)$ in optics. These lengths are determined by
comparing the QP $U_{qu}$ with the nonlinear term in the equation of
motion. Eq. (\ref{bogoliubov}) demonstrates that the propagation at
high frequencies corresponds to the so called {\em superluminal}
case. For $l_h = 0$, i.e. neglecting QP, $\Omega(k)$ reduces to a
linear dispersion law. It is interesting to indicate that the
Bogoliubov spectrum of excitations appears not only in the quantum
BEC system, for which it has been originally proposed, but also in
the case of classical electromagnetic wave propagating in the Kerr
medium.

Eq.(\ref{bogoliubov}) has four solutions, which for ${\bf v}_0 = 0$
read
\begin{equation}\label{bogoliubov-a}
k^2 = - \frac{1}{l_h^2}\left(1 \pm  \sqrt{1 + \frac{2 \omega^2
l_h^2}{\overline{s}\ ^2}}\right)
\end{equation}
Two of the four solutions ($-$ sign in Eq.(\ref{bogoliubov-a})) are
real and correspond to two plane waves moving in the mutually
opposite directions. The other two solutions ($+$ sign in
Eq.(\ref{bogoliubov-a})) are imaginary and correspond to excitations
growing or decaying exponentially on the scale $ l_h$. They are
usually of a minor physical importance. The situation may change for
the supersonic regime when four real solutions are possible,
provided $\omega < \omega_{max}$.\cite{X5a,X5b}

\subsection{Hydrodynamic approximation}

The hydrodynamic approximation, applied in many cases, assumes that
the QP $U_{qu}$ in Eq.(\ref{Euler}) is neglected. It is justified
when the QP is small as compared to the interaction energy, $U_{qu}
\ll g f_0^2$, i.e. on the large scales $|x| \gg l_h$. Now again it
is possible to eliminate the variable $\chi$ in Eq.
(\ref{linear-b}), which is then reduced to the equation
\begin{widetext}
\begin{equation}\label{linear-f}
\frac{\hbar^2}{gf_0^2}\left\{[\partial_t^2 + \partial_t {\bf v}_0
\cdot\nabla]\xi + \nabla(\cdot{\bf v}_0\partial_t\xi) +
\nabla(\cdot{\bf v}_0({\bf v}_0\cdot \nabla))\xi - \nabla( s^2
\nabla)\xi\right\} = 0.
\end{equation}
\end{widetext}
for the phase fluctuation field $\xi$.

This equation has the form of a Klein-Gordon equation in a curved
space, which has been discussed many times starting from the seminal
paper by Unruh\cite{U81} (see also Refs.
\onlinecite{BLV03,LKO03a,LKO03b,FFBF10}). In particular, we may
apply it to the situation, which takes place in the Laval nozzle
(Fig. \ref{Laval.fig}), creating a transonic flow of a luminous
fluid.\cite{FFBF10} In this case we may reduce the problem to 1+1
dimensions keeping the time coordinate and one spatial coordinate
$x$ along the streamline crossing the Mach horizon at $x = 0$. In
its narrow vicinity the velocity is a linear function of the
coordinate, $v_0(x) \approx \overline{s} (1 + \alpha x)$. Since the
density $\rho_0(x)v_0(x) = \frac{m^2}{g} s^2(x)v_0(x)$ of the flow
reaches its maximum at the throat of the nozzle we have $s(x)
\approx \overline{s} (1 - \frac{1}{2}\alpha x)$. Therefore the
functions $v(x)$ and $s(x)$ cannot be chosen independently unless we
allow for the $x$ dependence of the nonlinearity
coefficient.\cite{X5b} Then Eq.(\ref{linear-f}) becomes
\begin{equation}\label{linear-g}
\left\{\partial_\tau^2 + 2 (1 + \alpha x) \partial_\tau \partial_x +
\alpha \partial_\tau + 3 \alpha \partial_x(x\partial_x) \right\} \xi
= 0
\end{equation}
to within the linear terms in $x$, where the variable $\tau =
t\bar{s}$ is used. This equation has two solutions, which at $|x|
\to 0$ behave as
\begin{equation}\label{eigen-a}
\xi_1 \propto e^{\gamma_0\ln x - i \omega t}, \ \ \xi_2 \propto e^{i
kx - i\omega t}
\end{equation}
where $\gamma_0 = i\frac{2\omega}{3\overline{s}\alpha}$ and
$k(\omega)= \frac{\omega}{\overline{s}}\frac{\omega + i\overline{s}
\alpha}{2\omega - 3i\overline{s} \alpha}$. Correspondingly we have
also two eigenfunctions for the density fluctuations
\begin{equation}\label{eigen-b}
\chi_1 \propto e^{(\gamma_0 - 1) \ln x - i \omega t}, \ \ \chi_2
\propto e^{i kx - i\omega t}
\end{equation}

These two solutions \cite{LKO03a,FFBF10} are the former propagating
plane waves in the Bogoliubov spectrum strongly distorted by the
accelerating background flow. The solution $\xi_1$ corresponds to
the excitation "attempting" to propagate against the flow in the
vicinity of the Mach horizon, which cuts it into two parts moving in
the opposite directions from the horizon. The solution $\xi_2$ is a
fluctuation propagating with the flow and therefore have a shape
quite similar to that of a plane wave. In the high frequency limit
these modes propagate with the double sound velocity.

Two remarks are in order: The exponent $\gamma_0$ here differs from
the corresponding exponent in Ref. \onlinecite{FFBF10} by a factor
of two due to the misprint in that paper. This exponent, as well as
the Hawking temperature to be obtained below, differs also from the
exponent in Ref. \onlinecite{LKO03a,LKO03b} by a numerical factor
due to the fact that here we take into account the coordinate
dependence both of the velocity of the flow and of the local sound
velocity.

An important feature of these two solutions is that they behave
completely differently at $|x| \to 0$. The solution $\xi_2$ is a
smooth function in this region meaning that the corrections due to
the neglected QP, proportional to the second derivative of $\xi_2$
are small and the above approximation is quite sufficient.
Completely different situation takes place when dealing with the
solution $\xi_1$, which is singular in the limit $|x| \to 0$. The
corrections to $\xi_1$ due to the QP diverge at $|x| \to 0$ and a
more refined procedure is necessary in order to regularize the
behavior of $\xi_1$ in this limit.

\subsection{Regularization}

As indicated above the eigenfunctions $\xi_1$ and $\chi_1$, obtained
in the hydrodynamic approximation, (\ref{eigen-a}) and
(\ref{eigen-b}), have a singularity (branching point) on the horizon
at $x = 0$. This type of behavior is typical of the eigenfunctions
of the Klein-Gordon equation in the vicinity of the horizon of a
real black hole (see, e.g., Ref. \onlinecite{DL08} and references
therein) and leads to the celebrated Hawking radiation.
Regularization in this case requires knowing the physics at the
Planck length scale (see, e.g. Ref. \onlinecite{BD84}). In the
systems discussed here the part of Planck length is played by $l_h$,
which is the healing length for BEC or nonlinearity length for the
luminous liquid. In order to regularize $\chi_1,\ \xi_1$ near the
horizon we have to take into account the QP in Eq. (\ref{linear-b}).

We apply now the same assumptions concerning the coordinate
dependence of the velocity and density near the Laval nozzle throat
as in the previous subsection. Using the Laplace representation
$$
\chi(x,\tau) = e^{-i\nu\tau} \int_C dk \chi_k e^{ikx}
$$
where $\nu = \omega/\overline{s}$, and the contour $C$ is chosen
such that the integral converges,\cite{BH86} then Eqs.
(\ref{linear-b}) become
\begin{equation}\label{linear-k}
\left\{
\begin{array}{c}
(-i\omega + i\overline{s} k) \chi_k - \alpha \overline{s}
\partial_k (k \chi_k) + i\alpha k \displaystyle\frac{\hbar }{m}
\xi_k + \displaystyle\frac{\hbar}{m} k^2 \xi_k  = 0,
\\ \\
- \displaystyle\frac{\hbar^2}{4m} \left[i\alpha k + k^2\right]
\chi_k - m\overline{s}^2 (\chi_k - i\alpha \partial_k \chi_k) +
\hbar [(-i\omega + i\overline{s}k)\xi_k - \alpha \overline{s}
\partial_k (k\xi_k)] = 0.
\end{array}\right.
\end{equation}
where the terms $O(\alpha^2/k^2)$ have been omitted. This
approximation is equivalent to considering a region $\alpha |x| \ll
1$ near the horizon. The representation (\ref{linear-k}) is
especially convenient. It allows us to solve explicitly the first
equation,
\begin{equation}\label{xi}
\xi_k = \frac{m }{\hbar} \frac{(i\omega - i\overline{s} k) \chi_k +
\alpha \overline{s} \partial_k (k \chi_k)}{i\alpha k  + k^2},
\end{equation}
and substitute the result into the second equation. We then get
equation
\begin{equation}\label{difequ-a}
\partial_k \ln \chi_k = \frac{1}{i\alpha k (2\nu - 3 k - i\alpha)}
\left[- \displaystyle\frac{l_h^2}{2} (i\alpha k + k^2)^2 - k^2 +
(\nu - k)^2 \right]
\end{equation}
for the function $\chi_k$. Two comments are in order with respect to
the approach taken in Ref. \onlinecite{X8} where the Laplace
representation has also been used. Although Eq. (\ref{difequ-a}) has
the structure similar to that of Eq. (23) in Ref. \onlinecite{X8},
it is certainly different. Most important is that the r.h.s. of Eq.
\ref{difequ-a} has two poles instead of a single pole at $k = 0$
(Ref. \onlinecite{X8} uses $k = - is$). This difference stems from
neglecting one of the terms in equation (12) in Ref.
\onlinecite{X8}. However, as we will see below, it is crucial for
our analysis not to neglect this term. An attention is also drawn to
the fact that the r.h.s. of Eq. \ref{difequ-a} involves a cubic term
with purely imaginary coefficient.

Eq. (\ref{difequ-a}) can be readily solved to within a term
independent of $k$,
\begin{equation}\label{solution-a}
\ln \chi_k = \gamma_1 \ln(k) + \gamma_2\ln(k - \frac{2}{3} \nu -
\frac{i}{3}\alpha) + \Lambda(k,\nu)
\end{equation}
where
$$
\gamma_1 = \frac{1}{4} - \frac{i \nu}{2\alpha}, \ \ \gamma_2 = -
\frac{1}{4} - i \frac{1}{6\alpha}\nu  - \frac{4i}{81 \alpha} l_h^2
\nu^3 + \frac{14}{81} l_h^2 \nu^2
$$
and the $l_h$ dependent part is given by
\begin{equation}\label{solution-b}
\Lambda(k,\nu) = \frac{l_h^2}{\alpha} \left \{- \frac{i }{18} k^3 +
\frac{5}{36} \alpha k^2  - \frac{i}{18} \nu k^2 - \frac{2i}{27}
\nu^2 k + \frac{4}{27} \nu \alpha k \right \}.
\end{equation}

\begin{figure}[tbp!]
\begin{center}
\includegraphics[width=0.95\columnwidth]{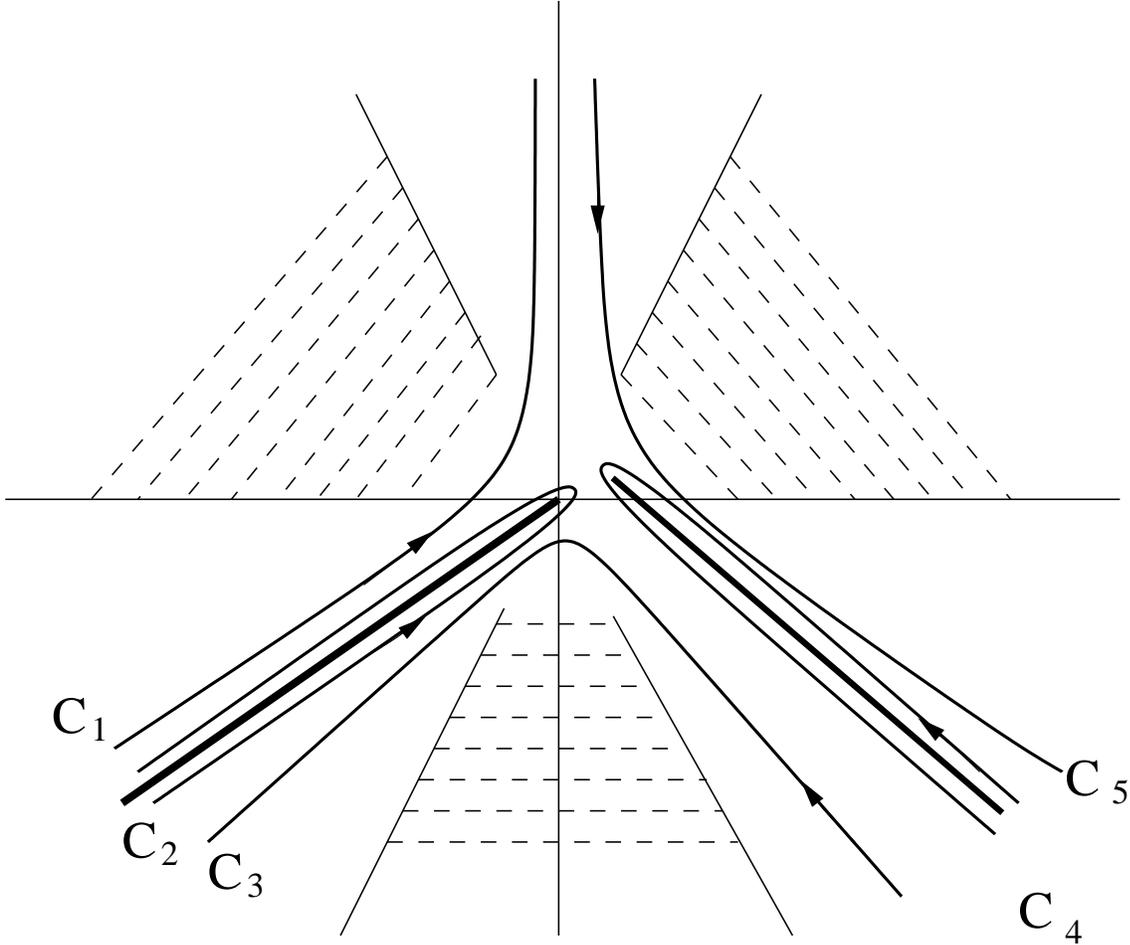}
\end{center}
\caption{Five possible contours in the complex $k$ plane for
calculation of the integral (\ref{integral}). The integral diverges
in the dashed sectors. The contours lie in the sectors of
convergency and also go around the two cuts, which start from the
two branching points at $k = 0$ and $k = \frac{2}{3}\nu +
\frac{i}{3} \alpha$.} \label{integral.fig}
\end{figure}

The real space solution of Eq. (\ref{linear-b}) is finally obtained
by the back Laplace transform
\begin{equation}\label{integral}
\chi(x,t) = e^{- i\nu\tau} \int_C dk k ^{\gamma_1} \left( k -
\frac{2}{3} \nu - \frac{i}{3}\alpha \right)^{\gamma_2} \exp\left\{
\Lambda(k,\nu) + ikx\right\}.
\end{equation}
This integral is calculated in the way similar to that used for the
calculation of Airy functions in Ref. \onlinecite{BH86}. First, we
have to find the sectors of convergency of this integral in the
Riemann plane for the complex variable $k$, which are controlled by
the term proportional to $k^3$ in Eq. (\ref{solution-b}). Hence the
condition Im$k^3 < 0$ must be required. It holds in the three
sectors: $2\pi/3
> \arg k > \pi/3$, $4\pi/3 > \arg k > \pi$ and $2\pi > \arg k > 5\pi/3$,
shown in Figure \ref{integral.fig}. The integrand's function has
also two branching points at $k= 0$ and $k = 2\nu/3 + i\alpha/3$,
which result in two cuts shown in Figure \ref{integral.fig}. Due to
these two cuts there are five possible integration contours $C_1$ to
$C_5$ and any other contour may be represented as a combination of
these five. They provide us with five different results for the
integral (\ref{integral}), however the sum of these five integrals
is zero due to the Cauchy theorem. As already mentioned above a
similar approach to the problem was applied in Ref. \onlinecite{X8}
with nearly the same configuration of contours as in Fig.
\ref{integral.fig}. However, the corresponding function in Ref.
\onlinecite{X8} has only one branching point. But to have four
independent solutions we need two branching points rather than one.
They are also important for Eq. (\ref{difequ-c}) to be derived
below.

The corresponding four integrals may be calculated by the steepest
decent method. Then each of these integrals will be a linear
combination of the contributions due to the four saddle points,
determined by the equation
\begin{equation}\label{saddle-a}
\partial_k \ln \chi_k = - ix.
\end{equation}
Alternatively we may also choose the integrals due to the individual
saddle points as linearly independent solutions. Substituting
$\partial_k\ln\chi_k$ from Eq. (\ref{difequ-a}) into Eq.
(\ref{saddle-a}) we get after some manipulations the $x$ dependent
dispersion relation
\begin{equation}\label{difequ-c}
\displaystyle\frac{l_h^2}{2} (i\alpha k + k^2)^2 \overline{s}^2 +
k^2 s^2(x) = \left[\omega - k v(x)\right]^2 = \Omega^2(k,\omega)
\end{equation}
where we remind that $v(x) \approx \overline{s}(1 + \alpha x)$,
$s(x) \approx \overline{s}(1 - \frac{1}{2} \alpha x)$ and $\omega =
\overline{s} \nu$. Here some terms of the order $O((\alpha/k)^2)$
and $O((\alpha x)^2)$ are kept only for the sake of a compact
presentation. However only the terms up to the first order are
meaningful.

Equation (\ref{difequ-c}) is rather similar to Eq. (6) in Ref.
\onlinecite{X8} and Eq. (23) in Ref. \onlinecite{X5b}. Nevertheless
we have to indicate important differences. Those equations relate to
the regions $\alpha|x| \gg 1$ far from the horizon where it is
assumed that the flow velocity reaches its supersonic or subsonic
limits, and the corresponding velocities, $v_\mp$, do not depend on
the coordinate $x$. Our equation (\ref{difequ-c}) is deduced for the
region close to the horizon $\alpha |x| < 0$ and both the flow and
sound velocities depend on the coordinate. It also contains an
imaginary correction in the quartic term.

\begin{figure}[tbp!]
\begin{center}
\includegraphics[width=0.4\columnwidth]{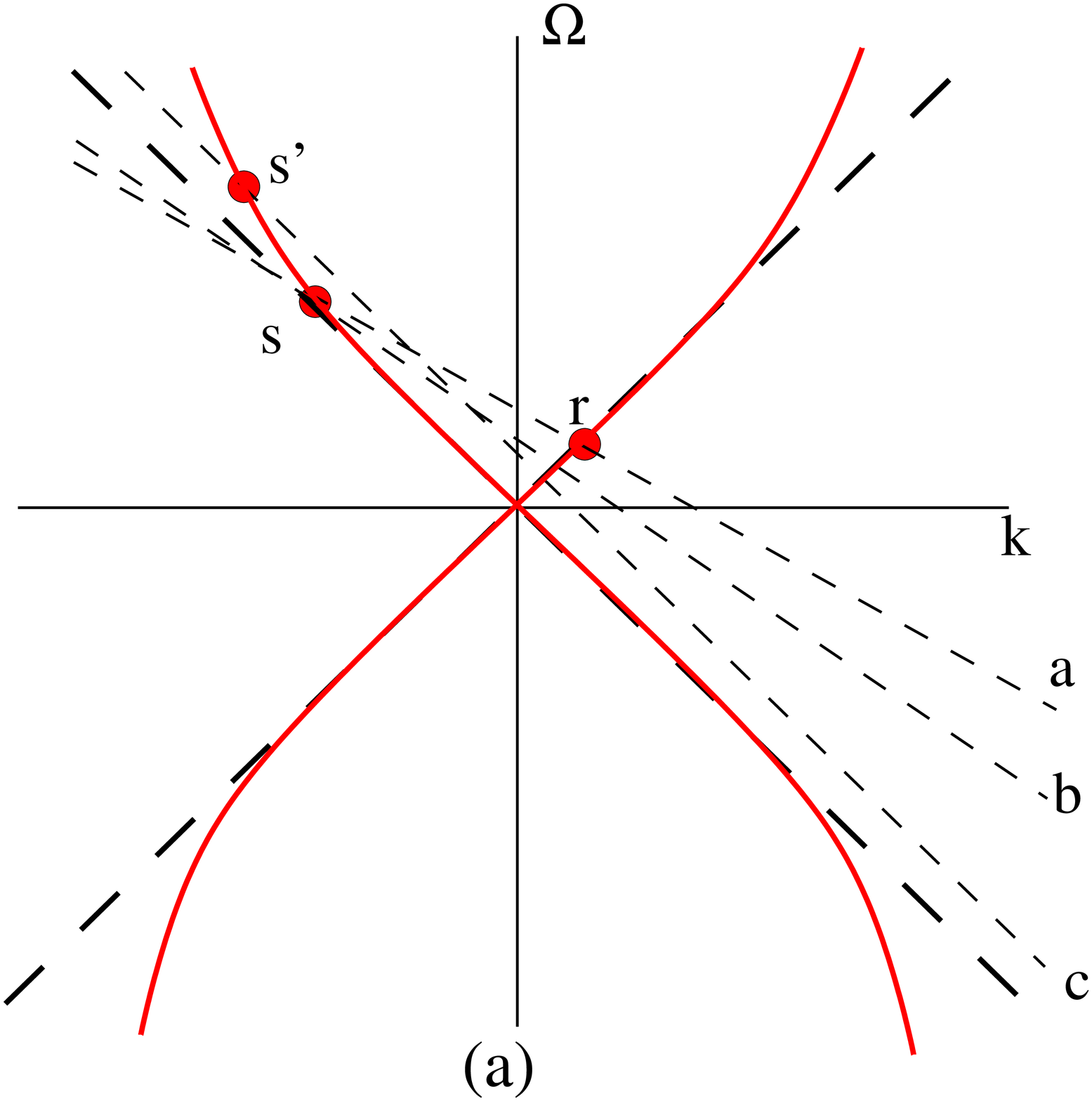}
\includegraphics[width=0.4\columnwidth]{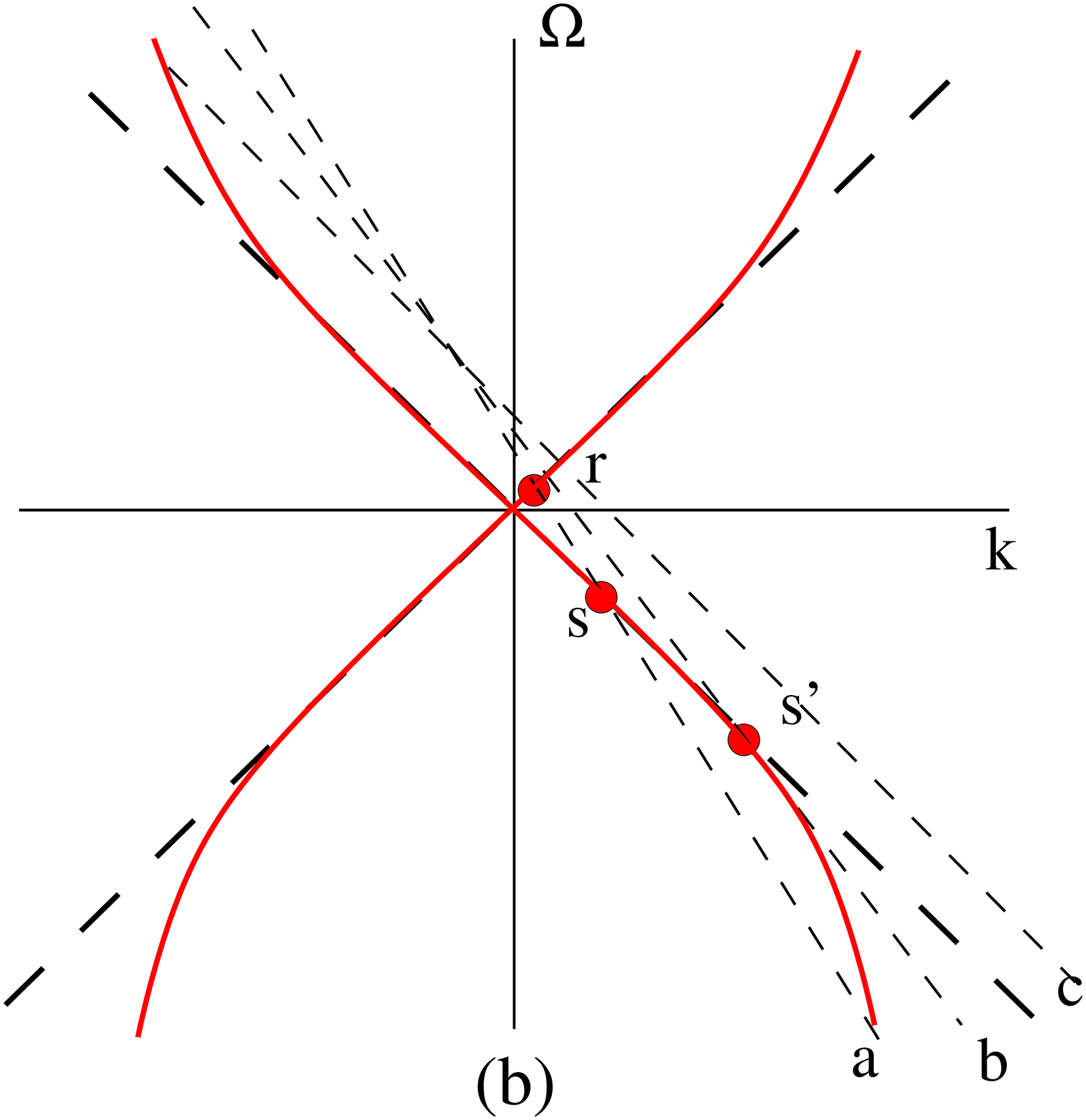}
\end{center}
\caption{(Color online) Graphical analysis of equation
(\ref{difequ-c}). Two branches of the function $\Omega(k,\omega)$
are plotted in red in figures (a) and (b). They are crossed by
straight dashed lines $\omega - kv(x)$ for three values of $x < 0$
((a) - subsonic regime) and three values of $x > 0$ ((b) -
supersonic regime). The full circles show the most important
solutions of the equation. The long dashed lines show the function
$\Omega(k,\omega)$ in the absence of the dispersion, i.e. $l_h =
0$.} \label{saddle.fig}
\end{figure}

An interesting insight may be obtained from this equation by a
graphical analysis, which is similar in some aspects to the one
carried out in Refs. \onlinecite{X7,X8} and particularly in Ref.
\onlinecite{X5b}. Let us neglect the imaginary correction to the
quartic term, which can be taken into account as a perturbation, and
plot both branches of the function $\Omega(k,\omega)$ crossed by the
three straight lines $\omega - k v(x)$ for three different values of
$|x|$, Fig. \ref{saddle.fig}. We also do not take into account the
$x$ dependence of the sound velocity, since it would have overloaded
the graphical presentation without changing any conclusions. All the
three lines are plotted for a fixed frequency $\nu$ but decreasing
distance $|x|$ from the horizon on both the subsonic side (Fig.
\ref{saddle.fig}a) and the supersonic side (Fig. \ref{saddle.fig}b).
The line (c) in both cases corresponds to $|x| = 0$ when $v(x) \to
\overline{s}$.

In the subsonic region (Fig. \ref{saddle.fig}a) where $x < 0$ we
always have two real solutions $r$ and $s$, with and without
dispersion. It is interesting to indicate that in the absence of the
dispersion, i.e in the absence of the quartic term in Eq.
(\ref{difequ-c}) when $l_h \to 0$ and when the two branches of
$\Omega(k,\omega)$ are just two straight lines with the slope one in
dimensionless units (long dashed lines in Fig. \ref{saddle.fig}),
the $s$ solution moves to infinity ($k_s \to - \infty$) as $x \to
0$. The dispersion changes this behavior so that $k_s$ tends to a
finite value in this limit (going from line (a) to (b) and then to
(c)) at the point (s') where the (c) straight line crosses the
$\Omega(k,\omega)$ curve.

In the supersonic case, the straight line (a) crosses the curve
$\Omega(k,\omega)$ in four points producing three right movers and
one left mover (not all are visible in Fig. \ref{saddle.fig}). It is
important to emphasize that two of these solutions, i.e. both right
movers ($r$) and ($s$) exist even in the absence of the dispersion.
If we move closer to the horizon, $|x| \to 0$, the slope of the
straight lines tends to that corresponding to the sound velocity
(line(c)). In the absence of the quartic term in Eq.
(\ref{difequ-c}) the $s$ solution moves to infinity as we approach
the horizon $x \to 0$.

Two solutions (both for subsonic and supersonic case)
$$
k_s = \frac{2\nu}{3\alpha x}, \ \ k_r =  \frac{2\nu}{4 - \alpha}.
$$
are obtained directly from Eq. (\ref{difequ-c}) at $l_h = 0$. The
saddle point $k_r$ leads us straightforwardly to the right mover
eigenfunction $\chi_2$ of Eq.(\ref{eigen-b}), which will be now
denoted by $\chi_r$. As for the $k_s$ point, the integral
(\ref{integral}) converges in a very broad range of $k$ values. We
may introduce the new variable $z = kx$ and write the integral
(\ref{integral}) in the approximate form
\begin{equation}\label{integral-a}
\chi_s(x,\tau) \approx x^{-(\gamma_1 + \gamma_2 + 1)}e^{-i\nu\tau}
\int dz z ^{\gamma_1 + \gamma_2} \exp\left\{ iz\right\}.
\end{equation}
where neglecting $\Lambda(k,\nu)$ in the exponent of the integrand
is justified as long as $x \gg l_h$. (A stricter inequality will be
given below). The validity of this equation is limited also by the
inequality $x \ll \min\{1/\nu,1/\alpha\} $. As a result we get the
singular eigenfunction
$$
\chi_s \propto x^{\gamma - 1},
$$
in which the parameter
\begin{equation}\label{exponent}
\gamma = - \gamma_1 - \gamma_2 = \frac{2i\nu}{3\alpha} +
\frac{4i}{81} \frac{l_h^2 \nu^3}{\alpha} - \frac{2}{27} l_h^2 \nu^2
\end{equation}
differs from the parameter $\gamma_0$ in Eqs. (\ref{eigen-a}) and
(\ref{eigen-b}) for $\xi_1$ and $\chi_1$, respectively, since it
contains small corrections due to the QP. That is why the notation
$\chi_s$ is used instead of $\chi_1$. This type of singular behavior
of the eigenfunction typically occurs near the horizon and is a
crucial ingredient in formation of the Hawking radiation ( see,
e.g., Refs. \onlinecite{DR76,DL08}).

Accounting for the quartic term in Eq. (\ref{difequ-c}) introduces
important changes. While moving towards the horizon, $|x| \to 0$,
for a given frequency we rotate the straight lines in Fig.
\ref{saddle.fig} so that the line $(a)$ becomes $(b)$ when two real
solutions merge (shown as the point s' in Fig. \ref{saddle.fig} b),
and then disappear (become complex) $(c)$. It actually means that
the singular behavior (\ref{integral-a}) of the eigenfunction
$\chi_s$ holds for not two small values of $|x|$ and becomes
regularized closer to the horizon. Here a remark is in order. It was
indicated in Refs. \onlinecite{X5a,X5b} that there is a maximal
frequency $\omega_{max}$, above which this pair of solutions does
not appear at all, which makes the Hawking radiation impossible.
This condition was obtained by considering the behavior of the
eigenfunctions in the asymptotic region $x \to \infty$ far from the
horizon where the flow velocity becomes constant. Our equation
(\ref{difequ-c}) allows for a smooth interpolation between these two
regions. It is clear that if we start at $\omega > \omega_{max}$ in
the supersonic asymptotic region those two solutions do not show up,
and then moving closer to the horizon, leading to a variation of the
sound and flow velocities, the singular solution $s$ will have no
chance to appear in the close vicinity of the horizon as well. If in
contrast $\omega < \omega_{max}$, both solutions exist for $x \to
\infty$. Now approaching the horizon a critical distance
$x_*(\omega)$ exists such that that these solutions disappear for $x
< x_*(\omega)$.

We may put the above pattern on a more quantitative basis. In order
to find the behavior of $\chi_s$ in the nearest vicinity of the
horizon we substitute $k_s \sim x^{-1}$ in the dimensionless
exponent $\Lambda$ in Eq.(\ref{integral}) and obtain that the cubic
term in Eq. (\ref{solution-b}) (former quartic term in
Eq.(\ref{difequ-a})) becomes large when
\begin{equation}\label{regularization}
|x| < l_r = \frac{ l_h}{( l_h\alpha)^{1/3}}.
\end{equation}
In this region the convergency of the integral (\ref{integral}) is
controlled by the cubic term in $\Lambda$ and the result remains
finite and not singular at $|x| \to 0$, i.e. the function $\chi_s$
is regularized at $|x| < l_r$. In order to demonstrate this it is
sufficient to calculate the integral
\begin{equation}\label{integral-b}
i\int_0^\infty z^{-\gamma} \exp(- \frac{1}{18} l_r^3 z^3 - zx) dz=
$$$$
= \frac{3}{l_r} \exp\left(\frac{i}{2} - \gamma\pi\right)
\left(-i\frac{l_r^3}{18}\right)^{-\frac{1}{3} - \gamma}
\left\{-2 \left(-i\frac{1}{18}\right)^{\frac{2}{3}}
\Gamma\left(\frac{1 - \gamma}{3}\right)
{}_1 F_2 \left(\frac{1 - \gamma}{3}; \frac{1}{3}, \frac{2}{3}; -
\frac{2x^3}{3l_r^3}\right)  +\right.
$$$$
2i \left(-i\frac{1}{18}\right)^{\frac{1}{3}} \frac{x}{l_r}
\Gamma\left(\frac{2 - \gamma}{3}\right) \
{}_1 F_2 \left(\frac{2 - \gamma}{3}; \frac{2}{3},
\frac{4}{3};-\frac{2x^3}{3l_r^3}\right)+
$$$$
\left. \frac{x^2}{l_r^2} \Gamma\left(\frac{3 - \gamma}{3}\right)
{}_1 F_2 \left(\frac{3 - \gamma}{3}; \frac{4}{3},
\frac{5}{3};-\frac{2x^3}{3l_r^3}\right)\right\}
\end{equation}
whose integrand is the approximate integrand of (\ref{integral}) at
large $k$. It represents the integral (\ref{integral}) over the
upper half of the contour $C_1$ or $C_5$. Here ${}_1F_2(a;b,c;x)$ is
a hypergeometric function and $\Gamma(x)$ is the Euler
Gamma-function. The other integrals corresponding to other contours
or their parts can be obtained by changing the relevant phases. This
integral tends to the finite limit
\begin{equation}\label{fluctuations-d}
\frac{1}{3} \left(-\frac{il_r^3}{18}\right)^{- \frac{1 - \gamma}{3}}
\exp\left(-\frac{i}{2} \gamma\pi\right) \Gamma\left(\frac{1 -
\gamma}{3}\right)
\end{equation}
for $x\to 0$, in fact for $|x| \ll l_r$. The derivatives with
respect to $x$ in this limit can be also calculated, resulting in a
regular Taylor expansion. One can also readily see that this
function may have zeros at complex $x$.

It is emphasized that the integral (\ref{integral-b}) depends on the
regularization length $l_r$ rather than on the healing length $l_h$.
It is also interesting that typically $l_h \alpha < 1$ so that
regularization starts at the scale larger than the healing length,
$l_r > l_h$.

\section{Hawking radiation from the Mach horizon.}

In order to calculate the frequency spectrum of the Hawking
radiation emanating from the Mach horizon we will apply the approach
similar to that proposed by Damour etal. \cite{DR76,DL08} The
central point in this approach is the calculation of the norm of a
straddled fluctuation, which clearly demonstrates how a negative
frequency state is cut into negative and positive frequency states
propagating in the opposite directions from the horizon. In our case
we have to calculate the norm (\ref{KGscalar}) for a system with
variation in the $x$ direction only:
\begin{equation}\label{norm}
<\vartheta_s,\vartheta_s> = \int dx \varrho_s(x) = -i\int dx
f_0^2(\chi_s^*\xi_s - \xi_s^*\chi_s)
\end{equation}
for the pair of eigenfunctions $(\xi_s,\chi_s)$. We have to deal
here with the two component function (\ref{two-component}), of which
$\chi_s$ calculated above is only one component. The second
component $\xi_s$ can be found now, say, by solving Eq.
(\ref{linear-b}). Neglecting again the contribution of the QP in Eq.
(\ref{linear-b}) we get
\begin{equation}\label{chi}
\chi_s = \frac{\hbar}{gf_0^2}\widehat{D}\xi_s
\end{equation}
so that
$$
\xi_s \propto x^{\gamma}
$$
for $\min\{1/\nu,1/\alpha\} \gg |x| \gg l_r$.

The two component field density and current density are
\begin{equation}\label{norm-a}
\begin{array}{c}
\varrho_s = - i \left\{[(\partial_t\xi_s^*)\xi_s -
\xi_s^*\partial_t\xi_s] + v_0(x) [(\partial_x \xi_s^*)\xi_s -
\xi_s^*\partial_x\xi_s]\right\},\\
j_s = i\left\{ v_0 (\xi_s^*\partial_t\xi_s -
(\partial_t\xi_s^*)\xi_s) + (v_0^2(x) -s^2(x))
(\xi_s^*\partial_x\xi_s - (\partial_x\xi_s^*)\xi_s)\right\}.
\end{array}
\end{equation}
where the factors $\hbar$ and $g$ are absorbed in the normalization
factor of $\xi_s$. The reader should also note that a second order
term of $\chi_s$, which is $O(l_h^4)$ has been neglected in $j_s$,
to be consistent with the neglect of $U_{qu}$. One can readily see
that the definition of the norm (\ref{norm}) coincides in this
approximation with the Klein-Gordon scalar product in the
corresponding curved space.

The local coordinate transformation
\begin{equation}\label{transformation}
d\widetilde{t} = dt + \frac{v_0(x)dx}{s^2(x) - v_0^2(x)}, \ \
d\widetilde{x} = dx
\end{equation}
proposed in Ref. \onlinecite{U81} can be also written as
$$
d\widetilde{x}^\mu = \left(
\begin{array}{c}
d\widetilde{t}\\
d\widetilde{x}
\end{array}
\right) = \Lambda^\mu_\nu dx^\nu.
$$
It allows us to rewrite the density and the current in the form
\begin{equation}\label{norm-c}
\begin{array}{c}
\widetilde{\varrho}_s = - i \displaystyle
\frac{s^2(\widetilde{x})}{s^2(\widetilde{x}) - v_0^2(\widetilde{x})}
[(\partial_{\widetilde{t}}\xi_s^*)\xi_s -
\xi_s^*\partial_{\widetilde{t}}\xi_s], \\
\widetilde{j}_s = - i[v_0^2(\widetilde{x}) - s^2(\widetilde{x})] [
(\partial_{\widetilde{x}} \xi_s^*)\xi_s -
\xi_s^*\partial_{\widetilde{x}} \xi_s],
\end{array}
\end{equation}
which have different signs in the subsonic $v(\widetilde{x}) <
s(\widetilde{x})$ and supersonic $v(\widetilde{x}) >
s(\widetilde{x})$ regions.

Now we use the density (\ref{norm-c}) in Eq. (\ref{norm}) in order
to calculate the norm. The corresponding integral is separated in
two regions outside the "black hole" $\widetilde{x} < -l_r$ (left)
and inside it $\widetilde{x} > l_r$ (right) (see Figure
\ref{Laval.fig}). We may neglect the contribution of the narrow
region $|x| < l_r$ in the integral where full solution of the second
Eq. (\ref{linear-b}), rather than Eq. (\ref{chi}), should be used.
As a result the norm takes the form
\begin{equation}\label{norm-b}
<\vartheta_s,\vartheta_s> \approx \int^{-l_r}_{-\infty}
d\widetilde{x} \widetilde{\varrho}_s + \int_{+l_r}^{\infty}
d\widetilde{x} \widetilde{\varrho}_s =
<\vartheta_s,\vartheta_s>_{left} +
<\vartheta_s,\vartheta_s>_{right}.
\end{equation}
The function $\widetilde{\varrho}_s$ in Eq. (\ref{norm-c}) diverges
at $|\widetilde{x}| \to 0$, which could have resulted in the
divergent contribution of the neglected region of integration.
However, it is important to emphasize that Eq. (\ref{norm-c}) holds
only outside the narrow region $|\widetilde{x}| > l_r$ near the Mach
horizon. Within this region, $|\widetilde{x}| < l_r$, we have to
return to Eq. (\ref{norm}). Although infinities are indicated as the
integration limits in Eq. (\ref{norm-b}), the principal
contributions come from the regions $\min\{1/\alpha, 1/\nu\} > |x| >
l_r$. Since both $\chi_s$ and $\xi_s$ are regular at $|x| < l_r$ we
will get only a small correction to the norm (\ref{norm-c}) from
this narrow region, which can be neglected as long as $l_h \ll
\min\{1/\alpha, 1/\nu\}$.

The integrals on the two sides of the horizon (left and right)
approximately obey the relation
$$
<\vartheta_s,\vartheta_s>_{left} = - e^{\displaystyle
2\pi\mbox{Im}\gamma} <\vartheta_s,\vartheta_s>_{right}
$$
due to the analytical properties of the function $\xi_s$. The total
negative norm (\ref{norm-b}) is cut into the left moving positive
frequency state and the right moving negative frequency state. The
left state propagates against the flow "outside the black hole". Its
relative weight is
\begin{equation}
N(\omega) = \left(e^{\displaystyle 2\pi\mbox{Im}\gamma} -
1\right)^{-1}.
\end{equation}
where
$$
\mbox{Im}\gamma = \frac{2\omega}{3\overline{s}\alpha} +
\frac{1}{162} \frac{\omega^3 l_r^3}{ \overline{s}\ ^3}
$$
contains a correction due to the QP proportional to the third power
of the frequency and, hence, $N(\omega)$ deviates from the Planck
distribution for the black body radiation. This result can be also
understood as a black body radiation spectrum with the frequency
dependent temperature
$$
T_H(\omega) = \frac{T_H(0)}{ 1 + \frac{1}{108} \frac{\omega^2 l_h^2
}{\overline{s}^2}}
$$
For not very high frequencies $\omega \ll \overline{s}/l_r$ this
dependence may be neglected and $N(\omega)$ becomes the standard
Planck distribution with the effective Hawking temperature
$$
T_H(0) = \frac{3 \hbar \overline{s} \alpha}{4\pi k_B}.
$$

\section{Conclusions}

The paper analyzes the role of the QP in the behavior of straddled
fluctuations in the nearest vicinity of the Mach horizon. These are
fluctuations attempting to propagate against the transonic flow that
rips them into two parts, one of which propagates against the
subsonic flow outside the "black hole", whereas the other part is
taken away by the supersonic flow and "falls down" inside the black
hole. The fluctuations are characterized by the singular behavior
(\ref{eigen-a}) in the region $l_r \ll x \ll
\min\{\overline{s}/\omega, 1/\alpha\}$, which on the quantum level
leads to the Hawking radiation from the horizon. The singularity at
the horizon $|x| \to 0$ is regularized due to the QP at the scale $x
< l_r$ (\ref{regularization}). This scale $l_r$ is usually larger
than the healing length $l_h$, which plays in this system the role
of the Planck length. Consequently  it is the length $l_r$ which is
relevant for the regularization rather than the "Planck length"
$l_h$ (healing length). This regularization length can also be
deduced from Eq. (57) of Ref. \onlinecite{X8}. However the author
did not make it explicit. The regularization removes the divergency
in the norm (\ref{norm-b}) and makes the procedure meaningful. The
influence of the QP can be felt also outside this scale at $|x| >
l_r$ since the exponent $\gamma$ in Eq.(\ref{exponent}), contains
now corrections due to the QP. They result in a deviation of the
high frequency tail of the spectrum of Hawking radiation from the
Planck black body radiation distribution.

Finally, we want to mention that neglecting QP results in a scale
invariance of the equation of motion for $|k| \to \infty$. This is
the origin of the occurrence of the singular solution $\chi_s
\propto x^{\gamma - 1}$ at $|x|\to 0$. The regularization we have
found close to the horizon after accounting for QP is based on
breaking this scale invariance for $|k| \to \infty$ (see Appendix)

{\bf Acknowledgment.} We are grateful to L. Frankfurt, N. Pavloff,
M. Reuter, M. Schwartz, G. Shlyapnikov and A. Soffer for helpful
discussions. VF is indebted to J. Gutenberg University, Mainz, where
he stayed during sabbatical, for hospitality. Support of United
States - Israel Binational Science Foundation, Grant N 2006242 is
acknowledged. RS also gratefully acknowledges financial support by
MPI-PKS, Dresden.

\appendix
\section{Function $\chi$ at $|x|\to 0$}

We try to elucidate here the mathematical features leading to the
singularity at the horizon and its regularization. Without
attempting mathematical rigor we present a qualitative discussion.
Let us start with the case where the QP is neglected. For a
stationary solution with variation in the $x$ direction only, we get
from Eq. (\ref{difequ-a})
$$
\partial_k \ln\chi_k = \frac{\nu^2 - 2\nu k}{i\alpha k (2 \nu - 3 k
- i \alpha)}
$$
for $l_h =0$ or for $|k| \to \infty$
\begin{equation}\label{App1}
\partial_k \ln\chi_k \approx - \gamma_0\frac{1}{k}
\end{equation}
If $\chi(x)$ is a generalized homogeneous function, i.e.
$$
\chi(\widetilde{x}) = \lambda^\gamma \chi(x), \ \ \ \widetilde{x} =
\lambda x
$$
for all real scaling parameters $\lambda$, then its Fourier
transform fulfills the condition
\begin{equation}\label{App2}
\chi_{\widetilde{k}} = \lambda^{(\gamma + 1)} \chi_k, \ \ \
\widetilde{k} = \lambda^{-1} k.
\end{equation}
Making use of (\ref{App2}) and $\partial_{\widetilde{k}} = \lambda
\partial_k$ it is easy to prove that
$$
\partial_{\widetilde{k}} \ln\chi_{\widetilde{k}} \approx -
\gamma_0\frac{1}{\widetilde{k}}
$$
i.e Eq. (\ref{App1}) is scale invariant. Taking the QP into account
Eq. (\ref{difequ-a}) results in
\begin{equation}\label{App3}
\partial_k \ln\chi_k \sim \frac{i}{6} l_r (l_r k)^2
\end{equation}
from which it follows that
$$
\partial_{\widetilde{k}} \ln\chi_{\widetilde{k}} \sim
\frac{1}{\lambda^3}\frac{i}{6} l_r (l_r k)^2
$$
i.e. Eq. (\ref{App3}) is not scale invariant. This is consistent
with the fact that its r.h.s., which involves the dimensionless
quantity $(l_r k)^2$, explicitly depends on the length scale $l_r$.
The latter has been found to be the scale below which regularization
occurs. Taking in the expansion of $v(x)$ and $s(x)$ (close to the
horizon) higher order terms in $x$ into account higher order
derivatives with respect to $k$ will occur in Eqs. (\ref{linear-k}).
This will prevent the decoupling of the fields $\xi_k$ and $\chi_k$
and therefore will hamper the analytical calculation of $\xi(x,t)$
and $\chi(x,t)$. Nevertheless the breaking of scale invariance still
holds and is generic. Consequently, the regularization close to the
horizon should be robust.

\end{document}